\documentclass{article}
\usepackage{ltwol2e}

\usepackage{psfig}

\def\Journal#1#2#3#4{{#1} {\bf #2}, #3 (#4)}

\def\NIM{\em Nucl. Instrum. Methods}

\def\NPB{{\em Nucl. Phys.} B}
\def\PLB{{\em Phys. Lett.}  B}
\def\PRL{\em Phys. Rev. Lett.}
\def\PRD{{\em Phys. Rev.} D}

\def\EPJ{{\em Eur. Phys. J.} C}

\def\be{\begin{equation}}
\def\ee{\end{equation}}
\def\bea{\begin{eqnarray}}
\def\eea{\end{eqnarray}}
\begin{document}
\title{Search for Neutral Heavy Leptons in the NuTeV Experiment at
       Fermilab}

\author{
 R.~B.~DRUCKER$^{6}$$^*$, T.~ADAMS$^{4}$, A.~ALTON$^{4}$, S.~AVVAKUMOV$^{7}$,
 L.~de~BARBARO$^{5}$, P.~de~BARBARO$^{7}$, R.~H.~BERNSTEIN$^{3}$,
 A.~BODEK$^{7}$, T.~BOLTON$^{4}$, J.~BRAU$^{6}$, D.~BUCHHOLZ$^{5}$,
 H.~BUDD$^{7}$, L.~BUGEL$^{3}$, J.~CONRAD$^{2}$, 
 R.~FREY$^{6}$, J.~FORMAGGIO$^{2}$, J.~GOLDMAN$^{4}$,
 M.~GONCHAROV$^{4}$,
 D.~A.~HARRIS$^{7}$, R.~A.~JOHNSON$^{1}$, S.~KOUTSOLIOTAS$^{2}$,
 J.~H.~KIM$^{2}$, M.~J.~LAMM$^{3}$, W.~MARSH$^{3}$, 
 D.~MASON$^{6}$, C.~McNULTY$^{2}$, K.~S.~McFARLAND$^{3,7}$, 
 D.~NAPLES$^{4}$, 
 P.~NIENABER$^{3}$, A.~ROMOSAN$^{2}$, 
 W.~K.~SAKUMOTO$^{7}$,
 H.~SCHELLMAN$^{5}$, M.~H.~SHAEVITZ$^{2}$, P.~SPENTZOURIS$^{2}$, 
 E.~G.~STERN$^{2}$, B.~TAMMINGA$^{2}$, M.~VAKILI$^{1}$, 
 A.~VAITAITIS$^{2}$, 
 V.~WU$^{1}$, U.~K.~YANG$^{7}$, J.~YU$^{3}$ and 
 G.~P.~ZELLER$^{5}$
}
\address{
 $^*$Presented by R.~B.~DRUCKER \\ 
 $^1$University of Cincinnati, Cincinnati, OH 45221 \\            
 $^2$Columbia University, New York, NY 10027 \\                   
 $^3$Fermi National Accelerator Laboratory, Batavia, IL 60510 \\  
 $^4$Kansas State University, Manhattan, KS 66506 \\              
 $^5$Northwestern University, Evanston, IL 60208 \\               
 $^6$University of Oregon, Eugene, OR 97403 \\                    
 $^7$University of Rochester, Rochester, NY 14627 \\              
}

\twocolumn[\maketitle\abstracts{
Preliminary results are presented from a search for neutral heavy
leptons in the NuTeV experiment at Fermilab.  The upgraded NuTeV
neutrino detector for the 1996-1997 run included an instrumented
decay region for the NHL search which, combined with the NuTeV
calorimeter, allows detection in several decay modes ($\mu
\mu \nu$, $\mu e \nu$, $\mu \pi$, $e \pi$, and $e e \nu$).  We see
no evidence for neutral heavy leptons in our current search in the
mass range from 0.3~GeV to 2.0~GeV decaying into final states
containing a muon.
}]

\pssilent

\section{Introduction}
\begin{figure}[hbt]
\centerline{\psfig{figure=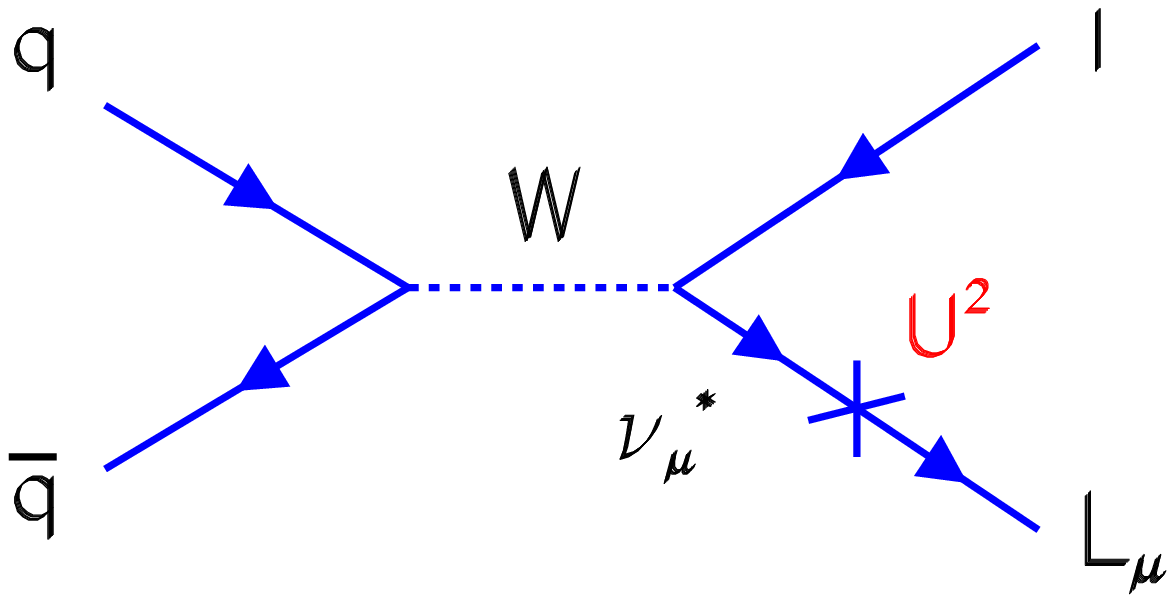,width=3.0in,bbllx=66pt,bblly=262pt,bburx=410pt,bbury=443pt}}
\centerline{\psfig{figure=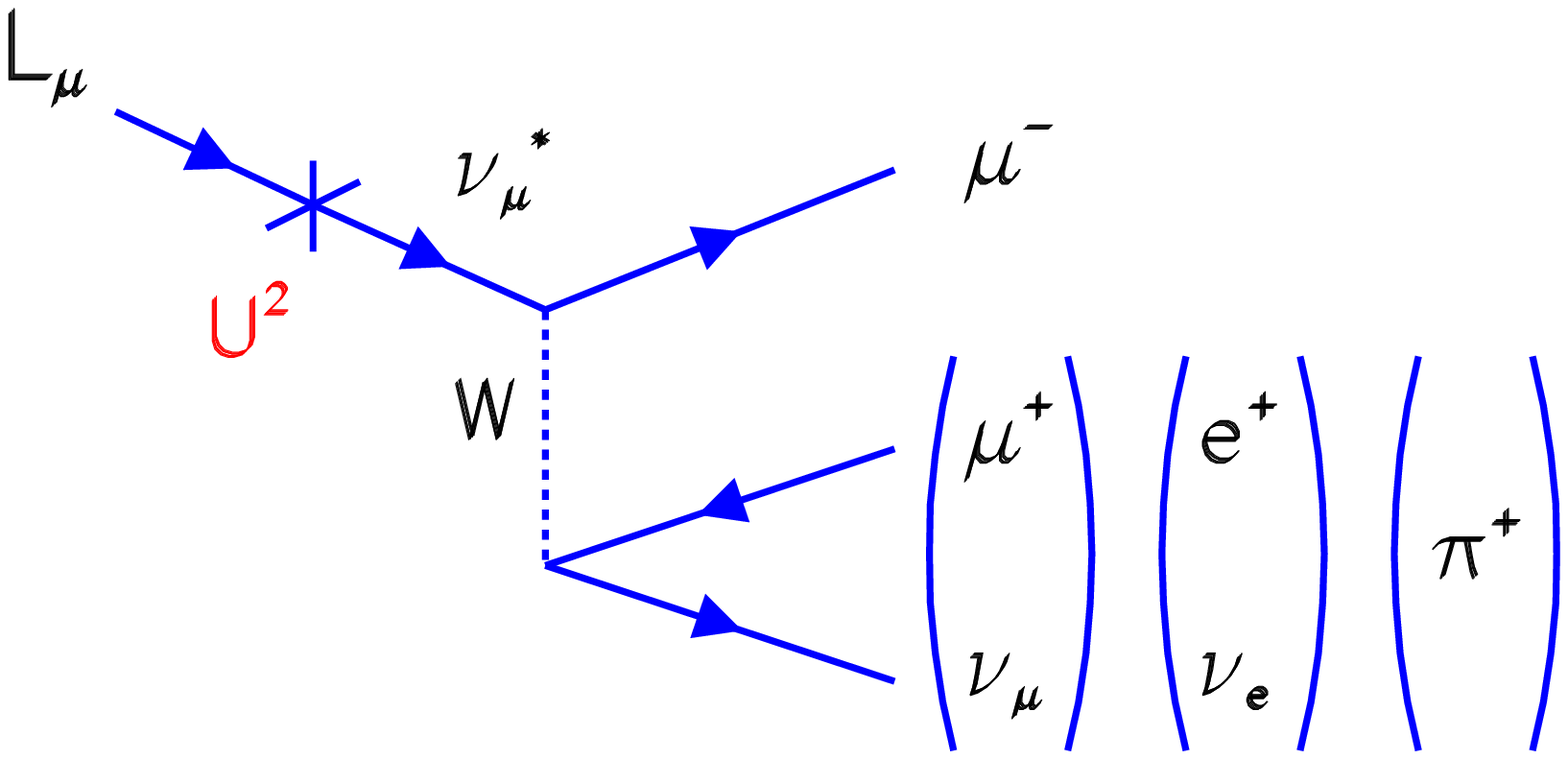,width=3.0in,bbllx=25pt,bblly=200pt,bburx=508pt,bbury=442pt}}
\caption[]{Feynman diagrams showing the production (from meson decay) and
           decay of neutral heavy leptons (L$_\mu$).  Decay via the
           Z$^{0}$ boson is also allowed, but not shown.}
\label{feynman}
\end{figure}

Many extensions to the Standard Model incorporating non-zero neutrino
mass predict the existence of neutral heavy leptons
(NHL).  See Refs.~[\ref{bib:grl}]~and~[\ref{bib:pdg}] for discussions
and references concerning massive neutrinos.
The model considered in this paper is that of Ref.~[\ref{bib:grl}] in
which the NHL is an iso-singlet particle that mixes with the Standard
Model neutrino.  Figure~\ref{feynman} shows the Feynman diagrams for
the production and decay of such an NHL.

The upgraded NuTeV detector includes a Decay Channel designed
specifically to search for NHL's and provides a significant
increase in sensitivity over previous searches.

\section{The Experiment}

The NuTeV calorimeter is described elsewhere~\cite{detector}; only
the features essential to this analysis are described here.  The
calorimeter consists of 84 layers of 10~cm steel plates and
scintillating oil counters.  A multi-wire gas drift chamber is positioned
at every 20~cm of iron for particle tracking and shower location.

\begin{figure*}[hbt]
\centerline{\psfig{figure=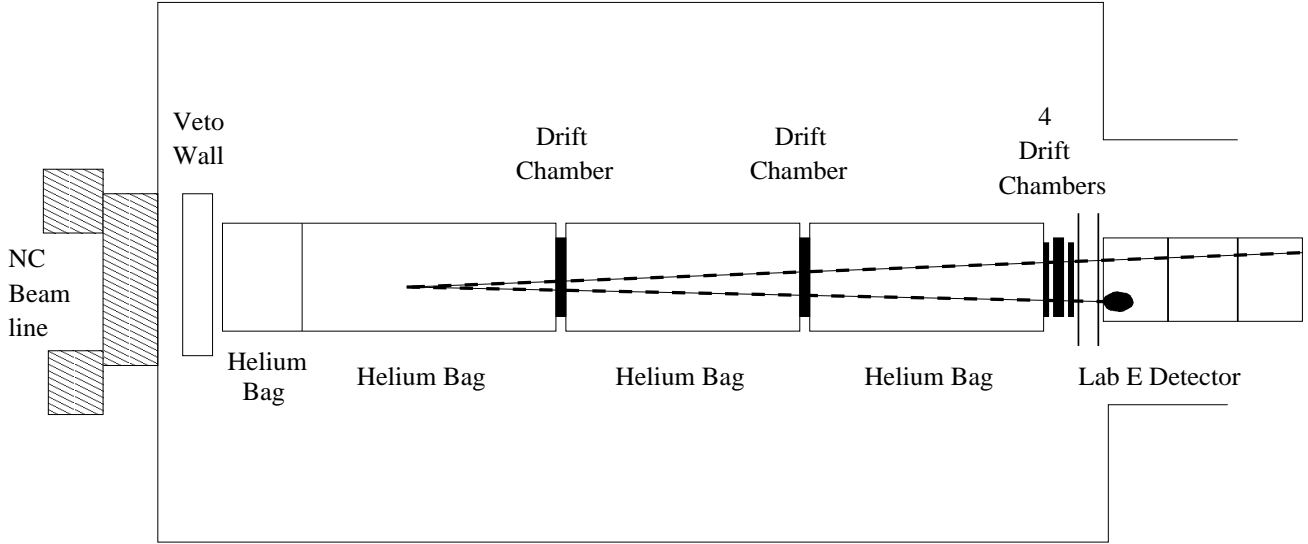,width=6.8in}}
\caption[]{A schematic diagram of the NuTeV decay channel.  The beam
           enters from the left, and at the far right is the NuTeV neutrino
           target.  An example of an NHL decay to $\mu \pi$ is also
           shown.  The event appears as two tracks in the decay channel,
           a long muon track in the calorimeter and a hadronic shower.}
\label{fig:dkchannel}
\end{figure*}

The decay channel is an instrumented decay space
upstream of the calorimeter.  The channel is 30~m long and filled with
helium using 4.6~m diameter plastic bags.  The helium was used to
reduce the number of neutrino interactions in the channel.  Drift
chambers are positioned at three stations in the decay channel
to track the NHL decay products.
Figure~\ref{fig:dkchannel} shows a schematic diagram of the decay
channel.  A 7~m $\times$ 7~m scintillating plastic ``veto wall'' was
constructed upstream of the decay channel in order to veto on any
charged particles entering the experiment.

\section{Event Selection}

Figure~\ref{fig:dkchannel} also shows an example of the event signature
for which we are searching.  The characteristics of an NHL event are
a neutral particle entering the channel and decaying in the helium region to
two charged (and possibly an additional neutral) particles.  The
charged particles must project to the calorimeter and at least one
must be identified as a muon.

To select events for this analysis we triggered on energy deposits of
at least 2.0~GeV in the calorimeter and required no veto wall signal.
We then require that there be
two well-reconstructed tracks in the decay channel that form a vertex
in the helium well away from the edges of the channel and the tracking
chambers.  The event vertex was required to be at least 3$\sigma$
away from the fiducial volume edges, where $\sigma$ is the resolution
of the vertex position measurement.  By requiring two tracks and
separation from the tracking chambers we greatly reduce the number
of background events from neutrinos interacting in the decay channel materials.
For all the cuts a vertex constrained fit is used in which the two tracks
are required to come from a single point in space.  The vertex resolution
depends on the opening angle of the tracks, but it is typically 25~cm
along on the beam axis and 2.5~cm transverse.

The two decay tracks are required to project to the calorimeter and to
match (in position) with particles identified in the calorimeter.
At least one of the two particles must be identified as a muon, because
for this analysis we only consider decay modes
with at least one muon.  In order to insure good particle
identification and energy measurement, we require all muons in the
event to have energy greater than 2.0~GeV and all electrons or hadrons
to have energy greater than 10.0~GeV.  These energy cuts also reduce
backgrounds from cosmic rays and neutrino interactions.

To further reduce acceptance for background events, additional kinematic
cuts are applied.  NHL decays are expected to have a small opening
angle; therefore, the decay particles are required to have slopes $p_x
/ p_z$ and $p_y / p_z$ less than 0.1 ($p_z$ is the momentum component
along the direction of the incoming beam, $p_x$ and $p_y$ are the
transverse components).  We are only considering NHL's produced by
kaon and charmed meson decays in this analysis; therefore, NHL's with
mass above 2.0~GeV are not considered.  We require the transverse
mass\footnote{The transverse mass is $p_T + \sqrt{p_T^2 + m_V^2}$,
where $p_T$ is the component of the total momentum of the two charged tracks
perpendicular to the beam direction
(i.e. the ``missing transverse momentum''), and $m_V$ is the invariant mass
of the two charged tracks.}
of the event to be less than 5.0~GeV in order to restrict ourselves
to this lower mass region.  Finally, in order to reduce neutrino-induced
events even further we form the quantities $x_{\rm eff}$ and $W_{\rm eff}$
by assuming that: i) the event is a neutrino charged current interaction 
($\nu N \rightarrow \mu N' X$), ii) that the highest energy muon comes 
from the neutrino-W vertex, and iii) the missing transverse momentum
in the event is carried by the final state nucleon.  We require
$x_{\rm eff} < 0.1$ and $W_{\rm eff} > 2.0$~GeV.

\section{NHL Monte Carlo}

\begin{figure}[hbt]
\centerline{\psfig{figure=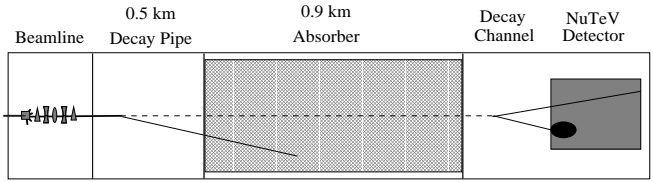,width=3.4in}}
\caption[]{A schematic diagram of the NuTeV beamline.  The 800~GeV proton
           beam from the Fermilab Tevatron enters from the left.
           NHL's are produced from the decays of kaons and pions in the
           decay pipe and from the decays of charm hadrons in the beam
           dumps.}
\label{fig:beamline}
\end{figure}

Figure~\ref{fig:beamline} shows a schematic of the NuTeV beamline.
The experiment took $2.5 \times 10^{18}$ 800~GeV protons from the 
Fermilab Tevatron on a BeO target.  Secondaries produced from the
target are focused in the decay pipe with a central momentum of
250~GeV.  The decay pipe is 0.5~km long, and the center of the
decay pipe is 1.5~km from the center of the decay channel.
Non-interacting protons, wrong-sign and neutral secondaries are
dumped into beam dumps just beyond the BeO target.
NHL's would be produced in decays of kaons and pions in the decay
pipe, as well as from charmed hadron decays in the primary proton
beam dumps.  Pion decays do not contribute significantly to this
analysis, as they cannot produce NHL's in the mass range of our
search.

\begin{figure}[hbt]
\centerline{\psfig{figure=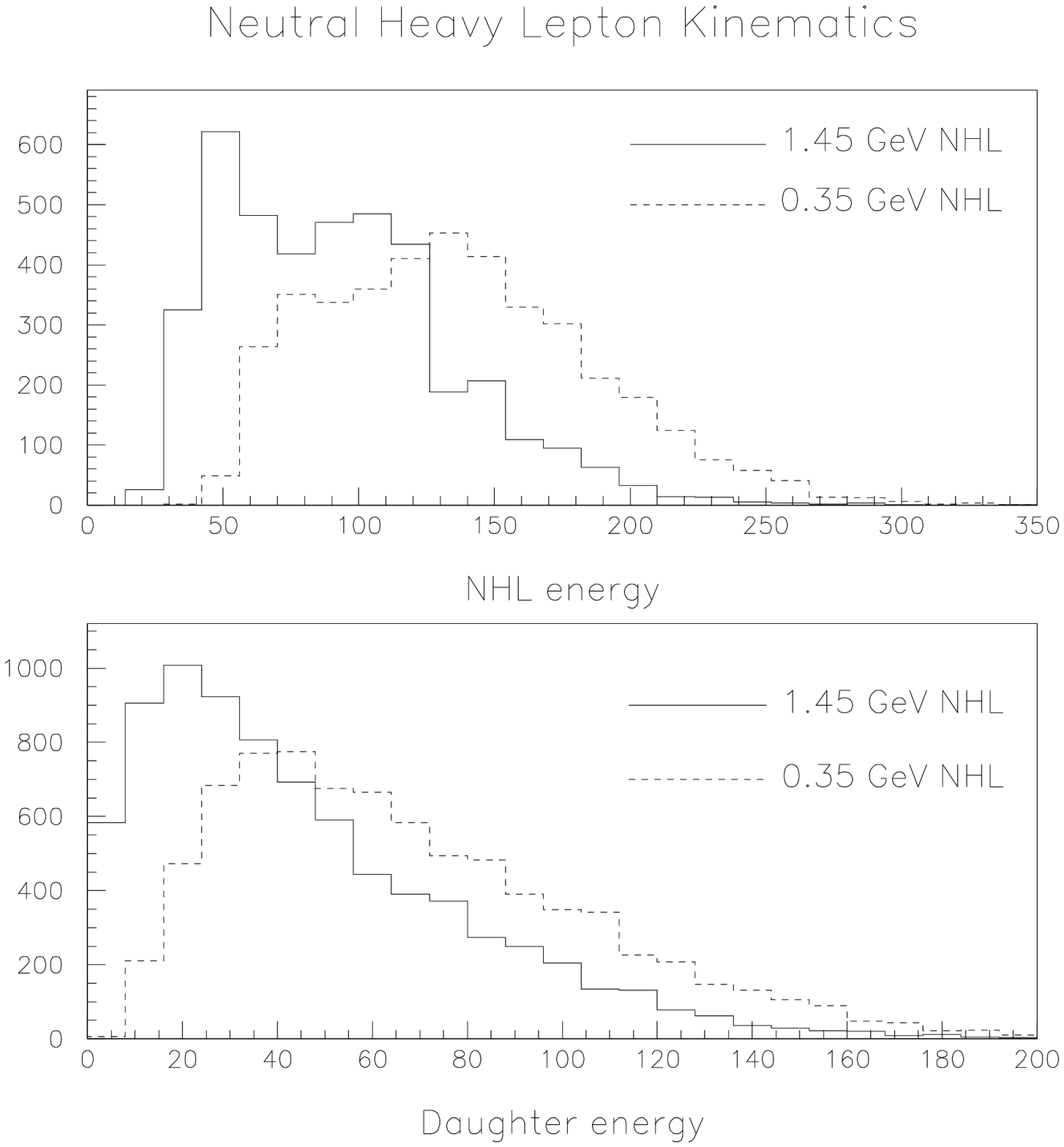,width=3.4in}}
\caption[]{The upper plot shows the energy distributions for Monte Carlo
           NHL's with mass 1.45~GeV and 0.35~GeV.  The lower plot shows
           the energy of the decay products of the NHL.}
\label{fig:nhlp}
\end{figure}

The production of kaons is simulated using the 
Decay Turtle~\cite{turtle} program.  The simulation of kaon decays 
to NHL's includes the effects of mass both in decay phase space and
in helicity suppression.  The production of charmed hadrons in the
beam dump are simulated using a Monte Carlo based on the production
cross sections reported in Ref.~[\ref{bib:charm}].  For this analysis
we only generate muon flavored NHL's.  Figure~\ref{fig:nhlp} shows
examples of the momentum distribution of NHL's produced by the NuTeV
beamline.  For a 1.45~GeV mass NHL, the average momentum is
$\sim$140~GeV.  For a 0.35~GeV mass NHL the average momentum is
$\sim$100~GeV.

The simulation of NHL decays uses the model of Ref.~[\ref{bib:tim}].  The
polarization of the NHL is also included in the decay matrix 
element~\cite{joe}.
The decay products of the NHL are run through a full Geant detector
simulation to produce simulated raw data which is then run through our
analysis software.

\section{Results}

We observe no events which pass our event selection cuts.
The number of expected background events are approximately 0.5.  The
largest background is 0.4 events expected from neutrino interactions
in the decay channel helium.  This estimate was made using the Lund
Monte Carlo~\cite{lund} to simulate neutrino--nucleon interactions.
In order to present a conservative
limit, we assume an expected background of zero events (this is only
a small change in the resulting limits).

\begin{figure}[hbt]
\centerline{\psfig{figure=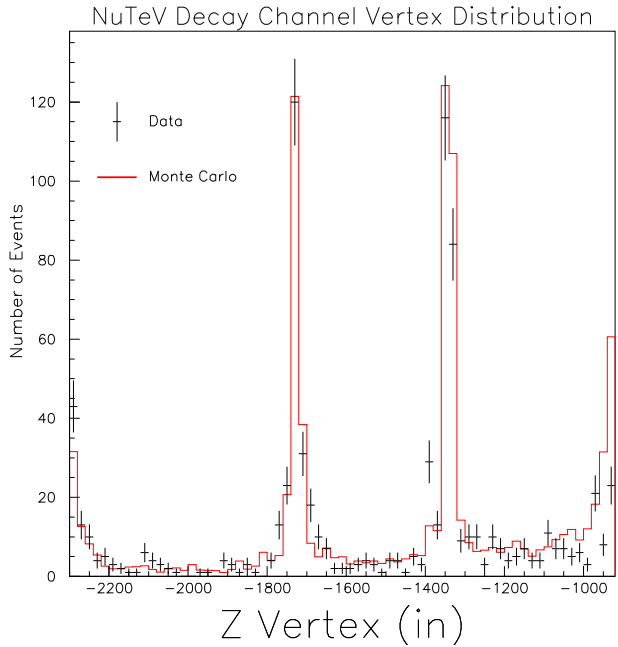,width=3.4in}}
\caption[]{The Z vertex distribution for neutrino interaction events in
           the NuTeV decay channel.  The points are data and the lines are
           Monte Carlo.  The peaks correspond to the positions of the
           drift chambers.}
\label{fig:zvert}
\end{figure}

In order to demonstrate the acceptance and reconstruction efficiency
of the experiment, we loosened several cuts in order to
examine the neutrino interactions in the decay channel material.  We
removed the cuts on the event vertex position (allowing events at the
positions of the chambers), and allow events with more than 2 tracks.
No calorimeter cuts (matching to particles, or energy cuts) were
applied, and no $x_{eff}$ or $W_{eff}$ cuts were applied.  
Figure~\ref{fig:zvert} shows the distribution of the event vertex
along the beam axis.  The peaks correspond to the positions of the
tracking chambers.  The plot also shows the neutrino
interactions in the helium gas between the chambers.  The number
of events seen is consistent with expectations.  This study demonstrates
that the channel and our tracking reconstruction are working well.

\begin{figure}[hbt]
\centerline{\psfig{figure=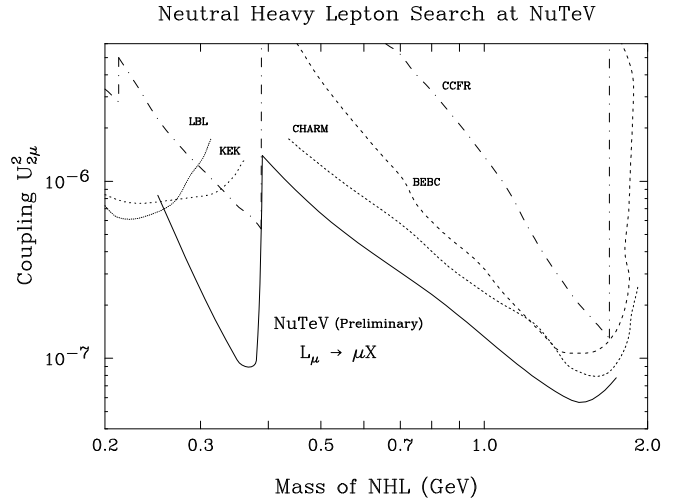,width=3.4in}}
\caption[]{Preliminary limits from NuTeV on the coupling, U$^2_{2\mu}$,
           of neutral heavy leptons (NHL) to the Standard Model left-handed
           muon neutrino as a function of NHL mass.  Only the $\mu$X decay
           modes of the NHL are included in this first search.  The limits
           are 90\% confidence and are based on zero observed events with
           zero expected background events.  The limits do not yet
           include effects from systematic uncertainties}
\label{fig:limits}
\end{figure}

Figure~\ref{fig:limits} shows our limits on the NHL--neutrino coupling,
$U_{2\mu}^2$, as a function of the mass of the NHL.  The results of previous
experiments~\cite{prev_ccfr,prev_bebc,prev_charm,prev_kek,prev_lbl} 
are shown for comparison.  Our
result is a significant increase in sensitivity in the range from 0.3~GeV
to 2.0~GeV.  These limits are for muon flavored NHL's and only
include their decay modes containing a muon.  The limits do not yet
include the effects of systematic uncertainties.

\section{Conclusions}

We have shown new preliminary limits from a search for muon flavored
neutral heavy leptons from the NuTeV experiment at Fermilab.  In
the future we plan to expand our search to include masses greater than
2.0~GeV as well as masses less than 0.3~GeV (perhaps to a final 
range of $\sim 0.020$~GeV to $\sim 10.0$~GeV).  We will also expand
our search to include electron flavored NHL's and all NHL decay modes
($\mu\mu \nu$, $\mu e \nu$, $\mu \pi$, $e \pi$, and $e e \nu$).

\section*{Acknowledgements}
This research was supported by the U.S. Department of Energy and the
National Science Foundation.  We would also like to thank the staff
of Fermliab for their substantial contributions to the construction
and support of this experiment during the 1996--97 fixed target run.

\end{document}